\documentclass[aps,pra,twocolumn,superscriptaddress,groupedaddress]{revtex4}  
\usepackage{graphicx}  
\usepackage{dcolumn}   
\usepackage{bm}        
\usepackage{amssymb}   

\usepackage{dsfont}
\usepackage{amsmath}
\usepackage{bbm}

\usepackage{here}
\usepackage{hyperref}
\usepackage{qcircuit}

\begin{document}

\newcommand{\ket}[1]{\left| #1 \right\rangle}
\newcommand{\bra}[1]{\left\langle #1 \right|}
\newcommand{\cA}{\mathcal{P}}
\newcommand{\cH}{\mathcal{H}}
\newcommand{\cL}{\mathcal{L}}
\newcommand{\cK}{\mathcal{K}}
\newcommand{\cM}{\mathcal{M}}
\newcommand{\cP}{\mathcal{P}}
\newcommand{\cR}{\mathcal{R}}
\newcommand{\cS}{\mathcal{S}}
\newcommand{\cU}{\mathcal{U}}

\newcommand{\bracket}[2]{\langle {#1}|{#2} \rangle}
\newcommand{\ketbra}[2]{\ket{#1}\negmedspace\bra{#2}}
\newcommand{\id}{\mathds{1}}


\title{Entanglement of two non-interacting qubits via a mesoscopic  system}
                            
\author{Maryam Sadat Mirkamali}
 \email{msmirkam@uwaterloo.ca}
\affiliation{Institute for Quantum Computing, University of Waterloo, Waterloo, Ontario  N2L 3G1, Canada}
\affiliation{Department of Physics and Astronomy, University of Waterloo, Waterloo, Ontario N2L 3G1, Canada}
\author{David G. Cory}%
\affiliation{Institute for Quantum Computing, University of Waterloo, Waterloo, Ontario N2L 3G1, Canada}
\affiliation{Department of Chemistry, University of Waterloo, Waterloo, Ontario  N2L 3G1, Canada}
\affiliation{Canadian Institute for Advanced Research, Toronto, Ontario  M5G 1Z8, Canada}
\affiliation{Perimeter Institute for Theoretical Physics, Waterloo, Ontario N2L 2Y5, Canada}
\author{Joseph Emerson}
\affiliation{Institute for Quantum Computing, University of Waterloo, Waterloo, Ontario N2L 3G1, Canada}
\affiliation{Department of Applied Mathematics, University of Waterloo, Waterloo, Ontario N2L 3G1, Canada}
\affiliation{Canadian Institute for Advanced Research, Toronto, Ontario  M5G 1Z8, Canada}

\date{\today}

\begin{abstract}
We propose a method for entangling two non-interacting qubits by measuring their parity indirectly through an intermediate mesoscopic system. The protocol is designed to require only global control and course-grained collective measurement of the mesoscopic system along with local interactions between the target qubits and mesoscopic system.
A generalization of the method measures the hamming weight of the qubits' state and probabilistically produces an entangled state by post-selecting on hamming weight one.
Our technique provides a new design element that can be integrated into quantum processor architectures and quantum measurement devices.
\end{abstract}

\maketitle


We show how a mesoscopic system (MS) can entangle two qubits by measuring the parity of the two qubits' wave-function.
The role of the MS is to magnify 
the qubits' parity such that the distinguishability of the two parity outcomes grows linearly with the MS's size. 
The initial state is separable over each qubit, and the qubits interact only with the MS. Relying only on collective control of the MS, a low-resolution collective measurement is still sufficient to prepare a post-selected entangled state with high confidence. 

A parity measurement is a two-outcome measurement that determines whether an even or odd number of qubits is in a particular logical state. 
For two qubits, each prepared in an equal superposition state, a projective measurement that reveals the qubits' parity but provides no information on individual qubits creates a post-selected entangled Bell state. 
Procedures for entangling two qubits through parity measurement have been proposed for different quantum systems \cite{Ruskov03, Trauzettel06, Williams08, Blais10, Tornberg10, Haack10} and performed experimentally with superconducting qubits \cite{Riste13, Saira14,Roch14,Chantasri16} and nuclear spins next to a nitrogen-vacancy center in diamond \cite{Pfaff13}. 

Here we propose implementing a projective parity measurement on two qubits indirectly through a MS. The MS in this model consists of identical two-level systems over which we have collective control.
This method leverages local interactions between each qubit and the MS and global control over the MS to correlate the two parity states of the qubits with the
distinguishable collective states of the MS. Global measurement of the MS and post-selection then creates an entangled state on the qubits due to the qubits' correlation with the MS. This indirect measurement must detect on the
order of $N$ excitations, where $N$ is the MS's size, 
unlike single qubit flip detection required in direct parity measurement of qubits. 


Previous proposals on entangling two qubits via an intermediate MS have used the MS to generate an effective interaction Hamiltonian between the target qubits \cite{Sorensen04,Trifunovic13}. 
In contrast, our novel approach relies solely on indirect joint measurement of the qubits facilitated by the MS.
By relying only upon the measurement, the distinguishability of the states of the MS corresponding to different parities of the qubits is a natural parameter for the success of our protocol. This distinguishability can be characterized over the classical probability distributions of the measurement outcomes. This characterization helps us derive a rigid upper bound for entanglement of the target qubits as a function of the MS's size and initial polarization. 
 Our analysis complements ongoing efforts to control the quantum aspects of MSs for quantum processing and metrology \cite{Cappellaro05, Appel09, Trifunovic13, Lukin01, Taylor03, Taylor032, Sorensen04}. 

In this work we introduce the general scheme for indirect parity measurement of two qubits through a MS. First, we present the general circuit
consisting of the evolution, measurement and post-processing steps and determine
the success criteria of the protocol.
Second, we explain the method with some idealized examples emphasizing on the role of the evolution step which magnifies the qubits' parity in the collective state of the MS. We show that collective control of the MS and local interaction between the qubits and the MS is enough to implement this magnification.
Next, we discuss the measurement step, demonstrating that a course-grained two outcome collective measurement on the MS with post-selection is sufficient for producing maximally entangled states on the qubits. Finally, we consider the effect of beginning with a non-ideal mixed initial state of the MS and find a rigid upper bound on the qubits' entanglement caused by MS's limited polarization.

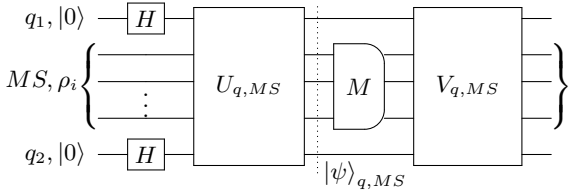
\begin{figure}[ht]
\centerline{
\Qcircuit @ C=1.2em @ R=0.2em {
\lstick {q_1, \ket{0}}  & \gate{H}&\multigate{8}{\rule{0.4em}{0em}U_{q,MS}\rule{0.4em}{0em}} \ar@{.}[]+<2.8em,0.5em>;[d]+<2.8em,-5.5em> & \qw & \multigate{8}{\rule{0.4em}{0em}V_{q,MS}\rule{0.4em}{0em}} & \qw \\
& & & & & \\
& \qw & \ghost{\rule{0.4em}{0em}U_{q,MS}\rule{0.4em}{0em}} & \multimeasureD {4}{M} & \ghost{\rule{0.4em}{0em}V_{q,MS}\rule{0.4em}{0em}}& \qw \\
\lstick {MS,  \rho_{i}\hspace{0.1cm}} & \qw & \ghost{\rule{0.4em}{0em}U_{q,MS} \rule{0.4em}{0em}} & \ghost{M} & \ghost{\rule{0.4em}{0em}V_{q,MS}\rule{0.4em}{0em}} & \qw \\
 & \vdots & & & &\\
 & & & & & \\
 & \qw & \ghost{\rule{0.4em}{0em}U_{q,MS}\rule{0.4em}{0em}} & \ghost{M} & \ghost{\rule{0.4em}{0em}V_{q,MS}\rule{0.4em}{0em}} & \qw \\
& & & & & & \\
\lstick {q_2, \ket{0}} &\gate{H}& \ghost{\rule{0.4em}{0em}U_{q,MS}\rule{0.4em}{0em}} & \qw  & \ghost{\rule{0.4em}{0em}V_{q,MS}\rule{0.4em}{0em}} & \qw
\gategroup{3}{1}{7}{1}{.8em}{\{}
\gategroup{3}{6}{7}{6}{.8em}{\}}\\
&&&\hspace{.5em}\ket{\psi}_{q,MS} &&
}
}
\caption{\label{fig:general} The general circuit of the parity measurement through an intermediate MS. 
}
\end{figure}

FIG. \ref{fig:general} shows a schematic of the proposed indirect parity measurement circuit.
Each qubit is prepared in the coherent $\ket{+}=\frac{1}{\sqrt{2}}(\ket{0}+\ket{1})$ 
state, and the MS is provisionally prepared in the polarized state, $\ket{0}^{\otimes N}$.
The MS evolves conditional on the qubits' state with the general unitary, $ U_{q,MS}=\ketbra{00}{00}_{q}\otimes U_{00}^{MS}+\ketbra{01}{01}_q\otimes U_{01}^{MS}+\ketbra{10}{10}_{q}\otimes U_{10}^{MS}+\ketbra{11}{11}_{q}\otimes U_{11}^{MS}$,
creating the following entangled state of the qubits and MS, 
\begin{eqnarray}
\ket{\psi}_{q,MS}&=&\frac{1}{2}(\ket{00}\otimes\ket{\psi_{00}}+\ket{01}\otimes\ket{\psi_{01}}  \nonumber \\
&+& \ket{10}\otimes\ket{\psi_{10}}+\ket{11}\otimes\ket{\psi_{11}}) 
\end{eqnarray}
with $\ket{\psi_{\gamma}}=U_{\gamma}^{MS}\ket{0}^{\otimes N}$ for $\gamma=00,01,10,11$.

The evolution is followed by a Positive-Operator Valued Measure (POVM) of the collective excitation on the MS and post-selection with the measurement operators $\{E_{\alpha}\}$ and the state-update-rule, $\rho_{MS;\alpha}=\frac{\sqrt{E_{\alpha}}.\rho_{MS}.\sqrt{E_{\alpha}}}{Tr(E_{\alpha}.\rho_{MS})}$. $\rho_{MS}$ and $\rho_{MS;\alpha}$ are the MS's states before and after the measurement with post-selecting the outcome $\alpha$, respectively \footnote{There is no general state-update-rule for the POVMs, and the post-measurement state depends on the details of the measurement procedure. Nevertheless any POVM is operationally equivalent to a von Neumann indirect measurement, which follows the mentioned state-update-rule. \cite{Joseph17}.}.
For a mesoscopic \textit{spin} system, this measurement corresponds to a total angular momentum measurement. 
 Adding the qubits, the state-update-rule becomes, 
\begin{eqnarray}
\label{eq:roqmssalpha}
\rho_{q,MS;\alpha}&=&\frac{(\id_2\otimes\sqrt{E_{\alpha}}).\rho_{q,MS}.(\id_2\otimes\sqrt{E_{\alpha}})}{Tr((\id_2\otimes E_{\alpha}).\rho_{q,MS})} \nonumber\\
 E_{\alpha}&=&\sum_{m=0}^{N}a_{\alpha,m}\Pi(m), \hspace{0.2cm} m=0,1,...,N
\end{eqnarray}
where $\Pi(m)$ is the operator that projects into the subspace with $m$ excitations and the coefficients $a_{\alpha,m}$ satisfy the conditions $0\leq a_{\alpha,m}\leq 1$ and $\sum_{\alpha}a_{\alpha,m}=1$, so that the POVM operators satisfy both the positivity, $E_{\alpha}\geq 0$,  and trace-preserving, $\sum_{\alpha}E_{\alpha}=\id$, conditions. 

\textit{To measure the target qubits' parity, the combination of the evolution and measurement must be such that the two pairs of states $\{\ket{\psi_{01}},\ket{\psi_{10}}\}$ and $\{\ket{\psi_{00}},\ket{\psi_{11}}\}$, called the odd and even pair, respectively, are discerned by the measurement but the states in each pair are not.
} 

With this criterion, the measurement and post-selection project the target qubits' state into even or odd parity subspaces due to its correlation with the  MS's state.
 If the MS's states within the odd and even pair are identical, post-selection on the measurement outcome ideally updates the state of the qubits to one of the two maximally entangled Bell states,  $\ket{e_{+}}:=\frac{1}{\sqrt{2}}\left(\ket{00}+ \ket{11}\right)$ or $\ket{o_{+}}:= \frac{1}{\sqrt{2}}\left(\ket{01}+\ket{10}\right)$, each with a probability of $\frac{1}{2}$. However, if the odd or even pair states are not identical, 
 a post-processing gate is required to disentangle the qubits from the MS (gate $V_{q,MS}$ in FIG. \ref{fig:general}). 
Thus, different and even orthogonal states in each pair are acceptable at the price of an extra gate after the measurement. 

Even if measuring the MS distinguishes between the even pair states, which corresponds to hamming weight measurement of the target qubits, there is a $\frac{1}{2}$ probability that the qubits will end up in the entangled state $\ket{o_{+}}$ with a hamming weight of one.  The remaining outcomes are hamming weights of zero and two, each with a probability of $\frac{1}{4}$ and with the updated states $\ket{00}$ and $\ket{11}$, respectively.  

We evaluate the success of our method by the amount of entanglement in the qubits' state, quantified by the fidelity, defined as the overlap of the qubits' state, $\rho_{q_1q_2}$, and the ideal maximally entangled state, $\ket{\phi}$, 
\begin{equation}
F_{\phi}(\rho_{q_1q_2}):=Tr(\rho_{q_1q_2}\ketbra{\phi}{\phi})=\bra{\phi}\rho_{q_1q_2}\ket{\phi}.
\end{equation}

The fidelity ranges between $0$ and $1$. If $F_{\phi}(\rho_{q_1q_2})>\frac{1}{2}$, the state $\rho_{q_1q_2}$ is entangled and can be distilled towards the maximally entangled state $\ket{\phi}$ \cite{Bennett96PRA,Bennett96PRL}. 

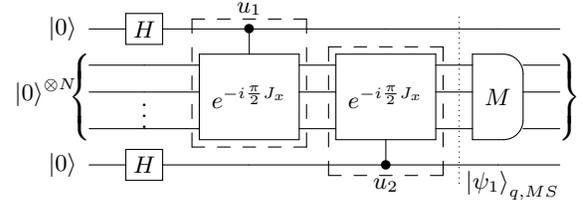
\begin{figure}[t!h]
      \centerline{
        \Qcircuit @ C=1.5em @ R=0.2em {
&  & u_1 &  & & \\
\lstick {\ket{0}}  & \gate{H}& \ctrl{2} & \qw  \ar@{.}[]+<3em,0.5em>;[d]+<3em,-5.5em> & \qw  & \qw \\
 &  & & & & \\
&  \qw & \multigate{4}{e^{-i\frac{\pi}{2} J_x}} & \multigate{4}{e^{-i\frac{\pi}{2} J_x}} & \multimeasureD {4}{M} & \qw \\
\lstick {\ket{0}^{\otimes N}} & \qw & \ghost{e^{-i\frac{\pi}{2} J_x}} &\ghost{e^{-i\frac{\pi}{2} J_x}}& \ghost{M}  & \qw \\
 & \vdots & & & &\\
& & & & &  \\
 & \qw & \ghost{e^{-i\frac{\pi}{2} J_x}} & \ghost{e^{-i\frac{\pi}{2} J_x}}& \ghost{M} & \qw \\
 & & &  & & \\
\lstick {\ket{0}} &\gate{H}& \qw & \ctrl{-2}& \qw   & \qw 
\gategroup{4}{1}{8}{1}{.8em}{\{}
\gategroup{4}{6}{8}{6}{.8em}{\}}
\gategroup{2}{3}{8}{3}{0.6em}{--}
\gategroup{4}{4}{10}{4}{.6em}{--} \\
&  & & u_2 & \hspace{1.2em}\ket{\psi_1}_{q,MS} &
}
       }     
\caption{\label{fig:example1}An idealized example of indirect parity measurement. During the evolution the MS's qubits are rotated conditioned on each external qubit's state, sequentially. The operator $J_x=\sum_j \sigma_{x}^{j}$  where $\sigma_{x}^{j}$ is the Pauli operator along x on the \textit{j}'th qubit. }
\end{figure}

FIG. \ref{fig:example1} shows an idealized parity measurement circuit. Two global $\pi$-rotations on the MS, conditioned on the state of each target qubit, correlate the target qubits' parity with the MS's collective excitation. 
The evolved state of the target qubits and MS is, 
\small
\begin{equation}
 \label{eq:PsiQMSE1}
 \ket{\psi_1}_{q,MS}=\frac{\ket{00}+\ket{11}}{2}\otimes\ket{0}^{\otimes N} 
+\frac{\ket{01}+\ket{10}}{2}\otimes\ket{1}^{\otimes N}.
 \end{equation}
 \normalsize
 The odd  pair states are equal to each other 
  and are therefore indistinguishable by any measurement, as are those of the even pair. Furthermore, the two pairs are maximally separated in the collective excitation spectrum, and thus can be distinguished with the lowest resolution global measurement.

In order to flawlessly distinguish between the states $\ket{0}^{\otimes N}$ with $m=0$ and $\ket{1}^{\otimes N}$ with $m=N$ number of excitations, it is sufficient that $Tr(E_{\alpha}.\Pi(0))\times Tr(E_{\alpha}.\Pi(N))=0$ for any measurement operator, $E_{\alpha}$.
If this condition
is not satisfied, the qubits' state will be perturbed from the maximally entangled states  $\ket{e_{+}}$ or $\ket{o_{+}}$. 
For example, if the measurement outcome is $\beta$, and the probabilities corresponding to even and odd pairs are  $p_{e,\beta}=Tr(E_{\beta}.\ketbra{0}{0}^{\otimes N})$ and $p_{o,\beta}=Tr(E_{\beta}.\ketbra{1}{1}^{\otimes N})$ where $p_{e,\beta}>p_{o,\beta}>0$, the qubits' entangled state is $\rho_{q_1q_2}=\frac{p_{e,\beta}}{p_{e,\beta}+p_{o,\beta}}\ketbra{e_{+}}{e_{+}}+\frac{p_{o,\beta}}{p_{e,\beta}+p_{o,\beta}}\ketbra{o_{+}}{o_{+}}$ with fidelity $F_{e_{+}}(\rho_{q_1q_2})=\frac{p_{e,\beta}}{p_{e,\beta}+p_{o,\beta}}$. 

A variation of the circuit outlined in FIG. \ref{fig:example1} measures the hamming weight of the target qubits and illustrates the role of the post-processing gate. If during the evolution step, half of the MS's qubits are flipped conditioned on the first qubit's state and the other half on the second qubit's state, the whole system evolves into the state, 
\begin{eqnarray}
\label{eq:PsiQMSE2}
\ket{\psi_2}_{q,MS}&=&\frac{1}{2}(\ket{00}\otimes\ket{0}^{\otimes N}+\ket{01}\otimes  \ket{0}^{\otimes \frac{N}{2}} \ket{1}^{\otimes \frac{N}{2}} \nonumber \\
&+&\ket{10}\otimes \ket{1}^{\otimes \frac{N}{2}} \ket{0}^{\otimes \frac{N}{2}}+\ket{11}\otimes\ket{1}^{\otimes N}).
\end{eqnarray}
Neither the odd nor even pair states are equal. However, the odd pair states share the same collective excitation, $m=\frac{N}{2}$, and thus are indistinguishable by any collective measurement. 
In contrast, the even pair states have the maximum separation in the collective excitation spectrum. 
 Nevertheless, depending on the details of the POVM on the MS, they may or may not be distinguished by the measurement,
 which correspond to indirect hamming weight and parity measurements of the qubits, respectively.
We illustrate the hamming weight measurement here and later discuss the parity measurement. 

Suppose that the POVM on the MS distinguishes between $m=0$, $m=\frac{N}{2}$ and $m=N$ number of excitations.
 Then, post-selection on the outcome $m=\frac{N}{2}$ projects the qubits' state into the odd parity subspace, $\frac{1}{\sqrt{2}}(\ket{01}\otimes  \ket{0}^{\otimes \frac{N}{2}} \ket{1}^{\otimes \frac{N}{2}}+\ket{10}\otimes \ket{1}^{\otimes \frac{N}{2}} \ket{0}^{\otimes \frac{N}{2}})$,  but the state of the qubits is entangled with the MS's state. Thus, a disentangling gate is required to restore the coherence of the qubits' state. A general choice for this gate is to reverse the evolution step, leading to the state $\frac{1}{\sqrt{2}}(\ket{01}+\ket{10})\otimes  \ket{0}^{\otimes N}$, which includes the maximally entangled $\ket{o_{+}}$ state on the qubits, separable from the MS's state. 

To perfectly distinguish the odd pair from both states of the even pair, any measurement operator, $E_{\beta}$, that can select the odd pair, i.e., $Tr(E_{\beta}\Pi(N/2))\neq 0$, must not overlap with the even pair, i.e.,  $Tr(E_{\beta}\Pi(0))=Tr(E_{\beta}\Pi(N))=0$. 
If this condition is not satisfied,
the fidelity of the updated state is $F_{o_{+}}(\rho_{q_1q_2})=\frac{2p_{01,\beta}}{2p_{01,\beta}+p_{00,\beta}+p_{11,\beta}}$ where $p_{\gamma,\beta}=Tr(E_{\beta}\ketbra{\psi_{\gamma}}{\psi_{\gamma}})$, with $\gamma=00,01,10,11$, is the probability of the measurement outcome $\beta$ corresponding to the MS's state $\ket{\psi_{\gamma}}$.

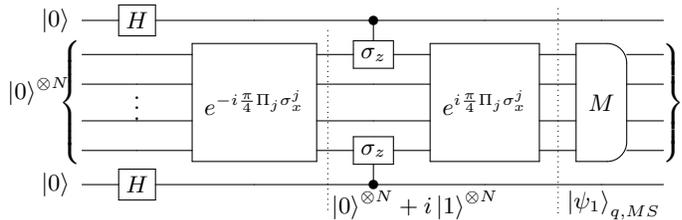
\begin{figure}[t!h]
    \centerline{
         \Qcircuit @ C=1.5em @ R=0.2em {
      \lstick{\ket{0}}   & \gate{H} & \qw & \ctrl{1} & \qw \ar@{.}[]+<3em,0.5em>;[d]+<3em,-6.5em> & \qw &\qw \\
        & \qw & \multigate{5}{e^{-i\frac{\pi}{4}\Pi_j \sigma_{x}^j}} \ar@{.}[]+<3em,1em>;[dd]+<3em,-4.5em>& \gate{\sigma_z} & \multigate{5}{e^{i\frac{\pi}{4}\Pi_j \sigma_{x}^j}}& \multimeasureD{5}{M} &\qw \\
        & \qw & \ghost{e^{-i\frac{\pi}{4}\Pi_j \sigma_{x}^j}}& \qw & \ghost{e^{i\frac{\pi}{4}\Pi_j \sigma_{x}^j}}& \ghost{M}   &\qw \\
       \lstick{\raisebox{1em}{$\ket{0}^{\otimes N}$}}  & \vdots & & & & & \\
        & & & & & & \\
        & \qw & \ghost{e^{-i\frac{\pi}{4}\Pi_j \sigma_{x}^j}}& \qw & \ghost{e^{i\frac{\pi}{4}\Pi_j \sigma_{x}^j}}& \ghost{M} &\qw \\
        & \qw & \ghost{e^{-i\frac{\pi}{4}\Pi_j \sigma_{x}^j}}& \gate{\sigma_z} & \ghost{e^{i\frac{\pi}{4}\Pi_j \sigma_{x}^j}}& \ghost{M} &\qw \\
    \lstick{\ket{0}} & \gate{H} & \qw & \ctrl{-1} & \qw & \qw & \qw         \gategroup{2}{1}{7}{1}{1 em}{\{} 
\gategroup{2}{7}{7}{7}{.8em}{\}}\\
    & & &\hspace{3.4em} \ket{0}^{\otimes N}+i  \ket{1}^{\otimes N}& & \hspace{1em}\ket{\psi_1}_{q,MS}  & 
        }     
        }       
        \caption{\label{fig:GHZ} Entanglement among the qubits in the MS and local interaction with target qubits. }
\end{figure}

The two examples discussed require a
collective rotation of the MS controlled by the target qubits' state. 
The target qubits' parity or hamming weight can also be encoded in the MS's collective state by the local interaction between each target qubit and its nearby qubit in the MS, by preparing the MS in an entangled state prior to their interaction.
FIG. \ref{fig:GHZ} shows this effect in its extreme limit. First the MS is prepared in the maximally entangled GHZ state, $\frac{1}{\sqrt{2}}(\ket{0}^{\otimes N}+i  \ket{1}^{\otimes N})$,
 by evolving under 
 the collective unitary operation $e^{-i\frac{\pi}{4}\Pi_j \sigma_{x}^j}$ where  $\sigma_{x}^j$ represents the Pauli operator along $x$ on the $j$th particle.
Second, the target qubits' parity is encoded in the phase of the GHZ state by applying two controlled-Z gates controlled by each target qubit on its nearby qubit in the MS. This global phase information is then transformed into population by
reversing the first gate, 
leading to the state $\ket{\psi_1}_{q,MS}$ in equation \ref{eq:PsiQMSE1},
the same as the evolved state of the first circuit. 

Similar to the first example, the qubits' parity is correlated with the states of the MS that are maximally separated in the collective excitation spectrum. This infers that these two circuits use the maximum capacity of the MS. 

\begin{figure}[t!h]
\centering
         \includegraphics[scale=0.5]{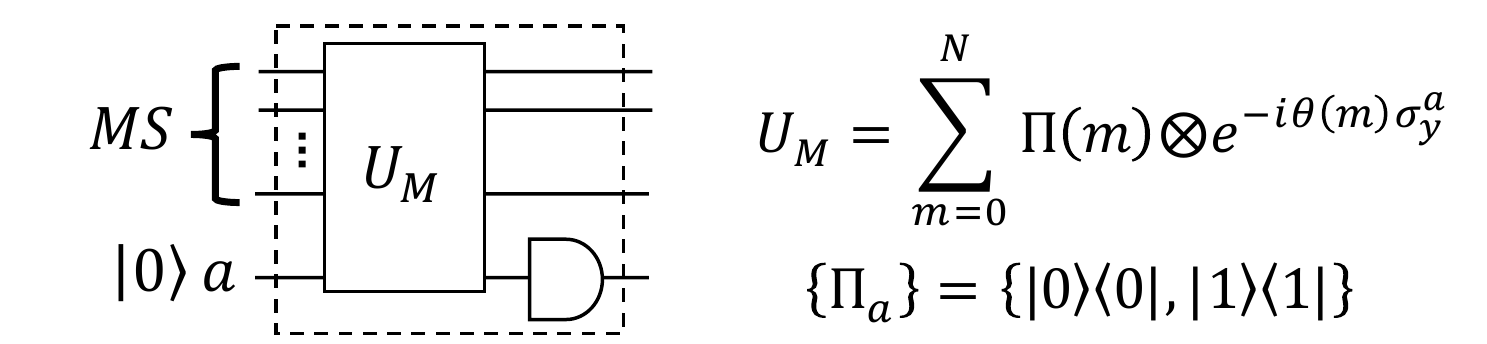}
                 \caption{\label{fig:measurement1} A simple two-outcome POVM for the MS implemented through collective interaction with an apparatus qubit.}    
\end{figure}

The evolution step is followed by a POVM of the MS. 
A wide range of measurements could achieve the desired goal. Here we show that to indirectly measure the qubits' parity,
a two-outcome POVM  that distinguishes between the odd and even pairs is sufficient.
Consider the general form of a two-outcome POVM of the collective excitation, parametrized with an angle $\theta(m)$, 
\begin{equation}
\label{eq:POVM}
E_0=\sum_m \cos\left(\theta(m)\right)^2 \Pi(m), \hspace{0.1cm} E_1=\sum_m \sin\left(\theta(m)\right)^2 \Pi(m)
\end{equation}	
Any two-outcome collective excitation measurement can be written in this form by properly choosing the corresponding function $\theta(m)$. According to Neumark's dilation theorem, any POVM on the system's Hilbert space, $H_S$, can be realized operationally as a projector valued measure (PVM) on an extended Hilbert space of the system and an apparatus $H_S\otimes H_A$ \cite{Neumark43}. This PVM can always be realized, operationally, as a von Neumann's indirect measurement \cite{Neumann32}. A von Neumann's indirect measurement consists of a unitary interaction between the system and the apparatus, followed by a PVM on the apparatus \cite{Joseph17}.
FIG. \ref{fig:measurement1} shows such an indirect measurement for the POVM in equation \ref{eq:POVM}. The gate $U_M=\sum_{m=0}^{N}\Pi(m)\otimes e^{-i\theta(m)\sigma_{y}^{a}}$
with $\Pi(m)$  and $\sigma_{y}^{a}$ acting on the MS and the apparatus qubit respectively,  rotates the apparatus qubit by an angle that depends on the MS's collective excitation. Next a PVM is performed on the apparatus qubit with the measurement operators, $\Pi_a(0)=\ketbra{0}{0}$ and  $\Pi_a(1)=\ketbra{1}{1}$. 
 The combination of the above unitary evolution and PVM effectively performs the POVM in equation \ref{eq:POVM} on the MS
 with the state-update-rule in equation \ref{eq:roqmssalpha}. 

The gate $U_M$ can be conveniently realized through collective linear interaction between the MS and the apparatus qubit with the Hamiltonian $H_M=g J_{z}\otimes\sigma_{y}^{a}$, where the operator $J_{z}$ is defined as $J_{z}=\sum_{j=1}^{N}(\sigma_{z}^j+\id^{j})/2=\sum_{m=0}^{N} m\Pi(m)$.
With this interaction, $\theta(m)=gmt_M$ is proportional to the collective excitation of the MS, $m$; where $g$ and $t_M$ are the interaction strength and time, respectively. 

\begin{figure}[t!h]
\centering
        \includegraphics[scale=0.5]{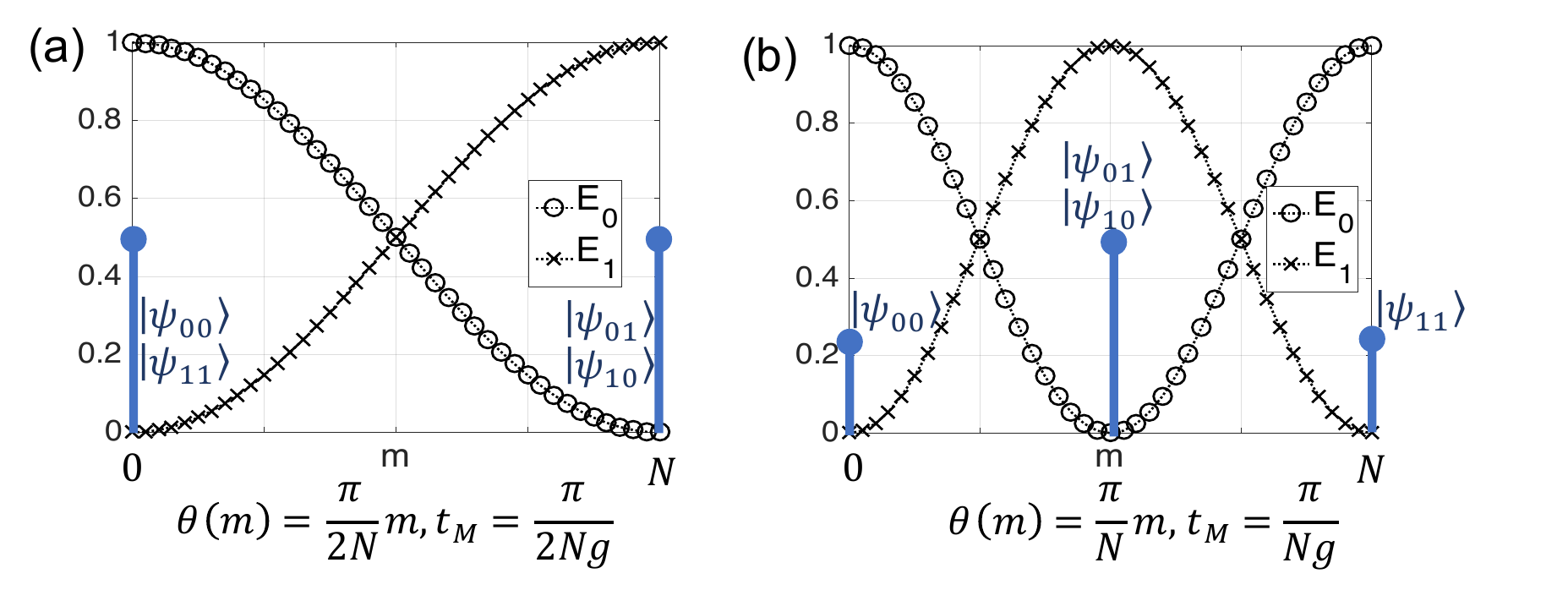}                \caption{\label{fig:measurement2} (a) The collective excitation spectrum of the MS with $\ket{\psi_{1}}_{q,MS}$ and the expansion of the appropriate two-outcome POVM. The measurement operator $E_0$ ($E_1$) selects the odd (even) pair over the even (odd) pair perfectly. (b) Similar to (a) but for the state $\ket{\psi_{2}}_{q,MS}$.
        }
\end{figure}

Based on the expected spectrum of the odd and even pair states prior to the measurement, $t_M$ is chosen to achieve the maximum contrast between the two pairs.  
FIG. \ref{fig:measurement2} shows the proper choices of $\theta(m)$ ($t_M$) for the examples discussed. 
For $\ket{\psi_1}_{q,MS}$, the two-outcome POVM with $\theta(m)=\frac{\pi}{2N}m$ ($t_M=\pi/2Ng$) flawlessly distinguishes between the two pairs,
resulting in the updated states $\ket{e_{+}}\otimes\ket{0}^{\otimes N}$ or $\ket{o_{+}}\otimes\ket{1}^{\otimes N}$, 
with equal probability.
For $\ket{\psi_2}_{q,MS}$,
the POVM with $\theta(m)=\frac{\pi}{N}m$ ($t_M=\pi/Ng$) 
updates the qubits-MS state to $\frac{1}{\sqrt{2}}(\ket{00}\otimes\ket{0}^{\otimes N}+\ket{11}\otimes\ket{1}^{\otimes N})$ or  $\frac{1}{\sqrt{2}}(\ket{01}\otimes\ket{0}^{\otimes N/2}\ket{1}^{\otimes N/2}+\ket{10}\otimes\ket{1}^{\otimes N/2}\ket{0}^{\otimes N/2})$, 
with equal probability.
These states will be evolved into the separable states between the qubits and the MS $\ket{e_{+}}\otimes\ket{0}^{\otimes N}$ and $\ket{o_{+}}\otimes\ket{0}^{\otimes N}$ by the following disentangling gate. With this POVM, the 
second example desirably measures the target qubits' parity, 
not their hamming weight,
since the two states $\ket{0}^{\otimes N}$ and $\ket{1}^{\otimes N}$ are not distinguishable by the measurement due to the cyclic form of the measurement operators expansion. 

 Next we discuss the mixed initial state of the MS; in particular, we find an upper bound on the qubits' entanglement when the MS is initially in the experimentally relevant mixed state,
\begin{equation}
\label{eq:ThMixed}
\rho_{\epsilon}=\left(\frac{\id+(1-\epsilon)\sigma_z}{2}\right)^{\otimes N}
\end{equation}
 where $N$ is the number of particles in the MS and  $(1-\epsilon)$ is the polarization of each.
 Without loss of generality we assume positive polarization; thus $0< (1-\epsilon) \leq 1$.
Consider the general indirect parity measurement circuit depicted in FIG. \ref{fig:general}.
In deriving the upper bound imposed by the initial state of the MS, we
allow any collective excitation measurement on the MS and any conditional unitary evolution of the form,
\small
\begin{equation}
\label{eq:untiaryP}
U_{q,MS}=(\ketbra{00}{00}+\ketbra{11}{11})\otimes V_e +(\ketbra{01}{01}+\ketbra{10}{10})\otimes V_o.
\end{equation}
\normalsize
The above unitary evolution guarantees that the odd pair states equal each other as do the even ones; therefore, a post-processing gate is not required. 
With a fixed evolution and measurement, the average fidelity over all measurement outcomes is,
 \begin{equation}
\label{eq:fidavg}
F_{avg}:=\sum_{\alpha} p_{\alpha} \mathcal{F}_{\alpha}(\rho_{q_1q_2,\alpha})
\end{equation}
where $p_{\alpha}$ is the probability of the measurement outcome $\alpha$. 
$\rho_{q_1q_2,\alpha}$ is the updated state of the qubits after the measurement on the MS with post-selecting the outcome $\alpha$, and 
$\mathcal{F}_{\alpha}(\rho_{q_1q_2,\alpha})$ is defined as, $\mathcal{F}_{\alpha}(\rho_{q_1q_2,\alpha}):=\max \left(F_{o_{+}}(\rho_{q_1q_2,\alpha}),F_{e_{+}}(\rho_{q_1q_2,\alpha})\right)$. 
This maximizing occurs as one of the two entangled states $\ket{o_{+}}$ or $\ket{e_{+}}$ is more probable depending on the measurement outcome. The corresponding fidelity is the appropriate measure of the entanglement. 

The average fidelity is related to the two classical probability distributions of the measurement outcomes corresponding to the states  $\rho_e=V_e\rho_{\epsilon}V_e^{\dagger}$ and  $\rho_o=V_o\rho_{\epsilon}V_o^{\dagger}$ of the MS measured by the collective POVM $\{E_{\alpha}\}$, as,
\begin{equation}
\label{eq:fidavg2}
F_{avg}(V_o,V_e,\{E_{\alpha}\})=\sum_{\alpha} \frac{1}{2} \max\left(p_{o,\alpha},p_{e,\alpha}\right)
\end{equation}
where $p_{o,\alpha}=Tr(\rho_o E_{\alpha})$ and $p_{e,\alpha}=Tr(\rho_e  E_{\alpha})$. See Appendix \ref{ap:1} for the derivation.
Maximizing the average fidelity over all pairs of unitary operators, $\{V_e, V_o\}$, and all collective POVMs, $\{E_{\alpha}\}$, gives the entanglement's upper bound,
\begin{equation}
\label{eq:fidAvgmax1}
F_{avg,max}:=\max_{V_o,V_e\{E_{\alpha}\}} F_{avg}(V_o,V_e,\{E_{\alpha}\}).
\end{equation}
This upper bound on the qubits' entanglement has an interesting physical interpretation as following. The classical trace distance of two probability distributions is defined as  $D_c(\vec{p}_1,\vec{p}_2):=\frac{1}{2}\sum_{\alpha}|p_{1,\alpha}-p_{2,\alpha}|=\sum_{\alpha} \max\left(p_{1,\alpha},p_{2,\alpha}\right)-1$, \cite{Joseph17, Nielsen00}. Thus the average fidelity can be written in terms of the classical trace distance between the probability distributions $\vec{p}_o$ and $\vec{p}_e$ as,
\begin{equation}
\label{eq:fidavg4}
 F_{avg}(V_o,V_e,\{E_{\alpha}\})=\frac{1}{2}\left(1+D_c(\vec{p}_o,\vec{p}_e)\right).
\end{equation}
The quantum trace distance between two states $\rho_1$ and $\rho_2$ 
is defined as the maximum of the classical trace distance between their associated probability distributions over all possible POVM measurements, $D_q(\rho_1,\rho_2):=\max_{\{E_{\alpha}\}} D_c(\vec{p}_1,\vec{p}_2)$. The quantum trace distance has an important operational meaning. It quantifies 
how distinguishable the two states are by a single shot measurement via the relation $\text{Pr}(\text{correctly inferring } \rho_1 \text{ over } \rho_2)= \frac{1}{2}(1+D_q(\rho_1,\rho_2))$ \cite{Joseph17,Nielsen00}.
According to equations \ref{eq:fidavg4} and \ref{eq:fidAvgmax1} the entanglement's upper bound is a function of the quantum trace distance of the states $\rho_o$ and $\rho_e$ as,
\begin{equation}
\label{eq:fidAvgmax2}
F_{avg,max}=\dfrac{1}{2}\left(1+\max_{V_o,V_e}D_q(\rho_o,\rho_e)\right).
\end{equation} 
Hence the desirable states, $\rho_o^{\star}=V_o^{\star}.\rho_{\epsilon}.V_o^{\star\dagger}$ and $\rho_e^{\star}=V_e^{\star}.\rho_{\epsilon}.V_e^{\star\dagger}$, that maximize the average fidelity are the ones that have the maximum quantum trace distance i.e. are the most distinguishable with a single shot measurement. Besides, the upper bound on the qubits' entanglement is the probability of successfully distinguishing between the states $\rho_o^{\star}$ and $\rho_e^{\star}$ of the MS with a single shot measurement, $F_{avg,max}=\frac{1}{2}\left(1+D_q(\rho_o^{\star},\rho_e^{\star})\right)$.

One choice for the states $\rho_e^{\star}$ and $\rho_o^{\star}$ are $\rho_e^{\star}=\rho_{\epsilon}=\left(\frac{\id+(1-\epsilon)\sigma_z}{2}\right)^{\otimes N}$ and $\rho_o^{\star}=\left(\frac{\id-(1-\epsilon)\sigma_z}{2}\right)^{\otimes N}$ corresponding to the gates $V_e^{\star}=\id^{\otimes N}$ and $V_o^{\star}=e^{-i \pi/2 J_x}=\sigma_x^{\otimes N}$;
 meaning that the best one can do is flipping all the two-level systems in the MS conditioned on the parity of the two target qubits similar to the circuit depicted in the FIG. \ref{fig:general}.
In this case, the optimal measurement on the MS is the PVM of the collective excitation, $\{E_{\alpha}^{\star}\}=\{\Pi(m)\}$
 \footnote{The coarse-grained PVM as long as $m\leq N/2$ and $m> N/2$ excitations do not mix are acceptable too, e.g., a two-outcome PVM with the  operators, $E_0=\sum_{m=0}^{\lfloor N/2 \rfloor} \Pi(m)$ and $E_1=\sum_{m=\lfloor N/2 \rfloor+1}^{N} \Pi(m)$.}. See Appendix \ref{sec:proof} for the proof.
 
Replacing  the above choices for $\rho_e^{\star}$, $\rho_o^{\star}$ and $\{E_{\alpha}^{\star}\}$ in the average fidelity relation in equation \ref{eq:fidavg2} gives the upper bound on entanglement as a function of the number of particles in the MS, $N$, and their polarization, $(1-\epsilon)$, 
\begin{eqnarray}
\label{eq:fidavgmax}
&&F_{avg,max}=\sum_{\alpha}\frac{1}{2}\max(b(\alpha;N,1-\frac{\epsilon}{2}),b(\alpha;N,\frac{\epsilon}{2}))  \nonumber \\
&& =\begin{cases}
B(\frac{N-1}{2};N,\frac{\epsilon}{2})  &\text{odd $N$} \\
B(\frac{N}{2}-1;N,\frac{\epsilon}{2})+\frac{1}{2}b(\frac{N}{2};N,\frac{\epsilon}{2}) &\text{even $N$}
\end{cases}
\end{eqnarray}
where the function $b$ represents the probability density function (PDF) and the function $B$ represents the cumulative distribution function (CDF) of the Binomial distribution. The second line follows the assumption that $\epsilon < 1$. 
Appendix \ref{sec:proof} provides a concrete general proof for equation \ref{eq:fidavgmax}.

\begin{figure}[t!h]
    \centering
        \includegraphics[scale=0.39]{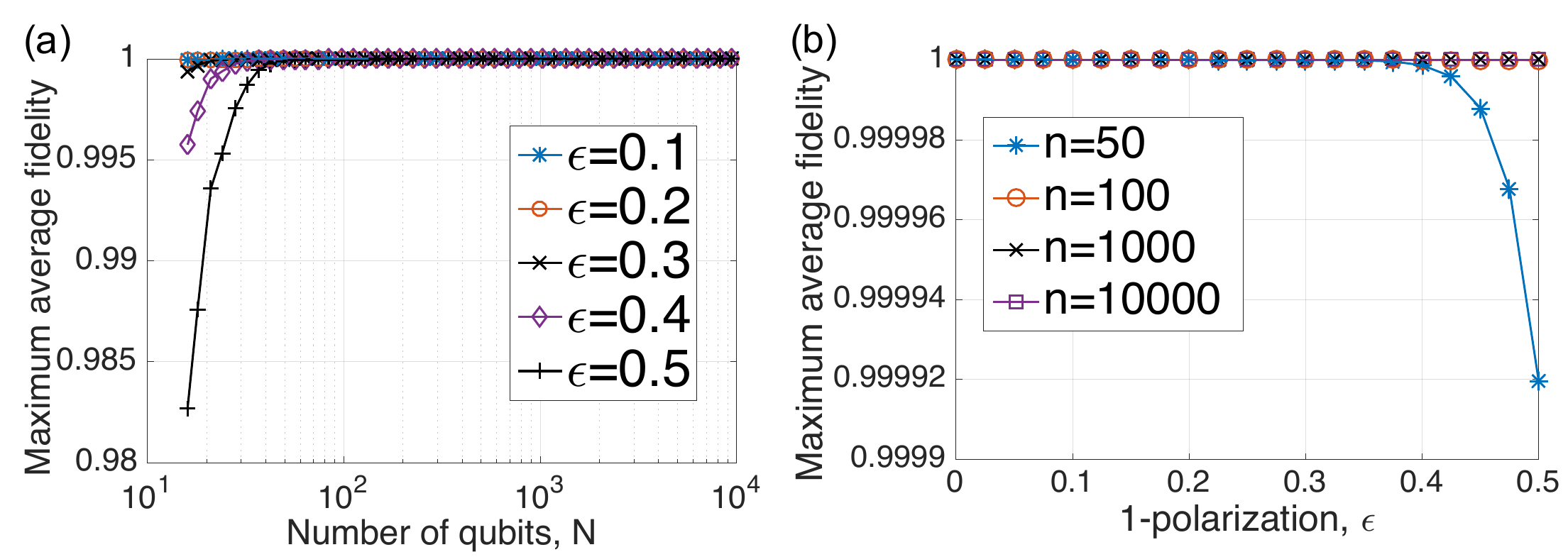}
        \caption{\label{fig:fid1} Upper bound on target qubits' entanglement (a) as a function of the number of qubits in the MS for different polarizations, (b) as a function of the polarization for different numbers of qubits in the MS .}
    \end{figure}

 FIG. \ref{fig:fid1} displays the plots of the entanglement's upper bound as a function of the number of qubits in the MS and their polarization. Increasing the number of qubits or average polarization raises $F_{avg,max}$, as expected. 
Moreover,
this bound is not limiting for physically realistic amounts of polarization e.g. with the polarization $1-\epsilon=0.5$ and only $50$ qubits the upper bound is remarkably $F_{avg,max}=0.9999$. 

In summary, we proposed a new procedure for entangling two non-interacting qubits in interaction with an intermediate mesoscopic system of identical two-level systems. The method is based on measuring the parity or hamming weight of the qubits' state indirectly by first coherently amplifying it in the collective state of the MS and then measuring the MS. 
This generic method is not limited to a specific MS or target qubit
 and is enabling in systems where measurement-device sensitivity is inadequate to detect a single qubit flip. 
 
We demonstrated that, 
by preparing the MS in an entangled state, the local interaction between target qubits and nearby qubit from the MS is enough to magnify the target qubits's parity in the collective state of the MS.
We discussed the measurement requirements and showed that with an ideal initial state  and evolution, a course-grained two-outcome collective measurement of the MS can result in maximally entangled states of the target qubits. We derived a rigid upper bound on entanglement of the target qubits imposed by the initial polarization of the MS's state and verified that our scheme performs well even under limited polarization. 
Other experimental limitations and imperfections such as available control and $T_1$ relaxation of the MS remain to be explored.

This work was supported by the Canada First Research Excellence Fund (CFREF),  the Canadian Excellence Research Chairs (CERC) program and the Natural Sciences and Engineering Research Council of Canada (NSERC) Discovery program.

\appendix
\section{Average fidelity as a function of classical probability distributions}
\label{ap:1}

Following the circuit \ref{fig:general} with the unitary evolution \ref{eq:untiaryP} the state of the target qubits and the MS after the evolution is:
\begin{eqnarray}
\label{eq:roqmssth}
\rho_{q,MS}&=&\dfrac{1}{2}\left(\ketbra{o_+}{o_+}\otimes \rho_o+\ketbra{e_+}{e_+}\otimes \rho_e \right. \\
&+&\left. \ketbra{o_+}{e_+}\otimes \chi_{oe}+\ketbra{e_+}{o_+}\otimes \chi_{eo}  \right) \nonumber
\end{eqnarray}
where $\rho_o=V_o . \rho_{\epsilon}.V_o^{\dagger}, \hspace{0.1cm} \rho_e=V_e . \rho_{\epsilon}.V_e^{\dagger},  \hspace{0.1cm} \chi_{oe}=V_o . \rho_{\epsilon}.V_e^{\dagger}, \hspace{0.1cm}\chi_{eo}=\chi_{oe}^{\dagger}=V_e . \rho_{\epsilon}.V_o^{\dagger}$. 
According to the state update rule \ref{eq:roqmssalpha}, the state of the qubits after the measurement, post-selection on the outcome $\alpha$, and tracing over the MS is:
\begin{eqnarray}
\rho_{q_1q_2,\alpha}&=&\dfrac{\ketbra{o_+}{o_+}Tr(E_{\alpha}\rho_o)+\ketbra{e_+}{e_+}Tr(E_{\alpha}\rho_e)}{Tr(E_{\alpha}\rho_o)+Tr(E_{\alpha}\rho_e)} \nonumber \\
&+&\dfrac{\ketbra{o_+}{e_+}Tr(E_{\alpha}\chi_{oe})+\ketbra{e_+}{o_+}Tr(E_{\alpha}\chi_{eo}}{Tr(E_{\alpha}\rho_o)+Tr(E_{\alpha}\rho_e)} \nonumber \\
&:=& \dfrac{p_{o,\alpha}\ketbra{o_+}{o_+}+p_{e,\alpha}\ketbra{e_+}{e_+}+ ...}{p_{o,\alpha}+p_{e,\alpha}}
\end{eqnarray}
where $p_{o,\alpha}:=Tr(E_{\alpha}\rho_o)$ and $p_{e,\alpha}:=Tr(E_{\alpha}\rho_e)$ are the probabilities of the measurement outcome $\alpha$ corresponding to the states $\rho_o$ and $\rho_e$ of the MS respectively.
The fidelity of the state $\rho_{q_1q_2,\alpha}$ with the odd and even parity states is,
\begin{eqnarray}
 \label{eq:fidelity12}
F_{o_+}(\rho_{q_1q_2,\alpha})&=&
Tr(\rho_{q_1q_2,\alpha}\ketbra{o_{+}}{o_{+}})=\dfrac{p_{o,\alpha}}{p_{o,\alpha}+p_{e,\alpha}} \nonumber\\
F_{e_+}({\rho_{q_1q_2,\alpha}})&=&Tr(\rho_{q_1q_2,\alpha}\ketbra{e_{+}}{e_{+}})=\dfrac{p_{e,\alpha}}{p_{o,\alpha}+p_{e,\alpha}}.
\end{eqnarray}
\normalsize
Substituting in equation \ref{eq:fidavg} we find the average fidelity for a particular choice of the evolution gates, $\{V_o,V_e\}$,  and the measurement, $\{E_{\alpha}\}$, 
\small
\begin{eqnarray}
\label{eq:fidavg3}
F_{avg}(V_o,V_e,\{E_{\alpha}\})&=&\sum_{\alpha} p_{\alpha} \max\left(\dfrac{p_{o,\alpha}}{p_{o,\alpha}+p_{e,\alpha}},\dfrac{p_{e,\alpha}}{p_{o,\alpha}+p_{e,\alpha}}\right) \nonumber \\
 &=&\sum_{\alpha} \frac{1}{2} \max\left(p_{o,\alpha},p_{e,\alpha}\right).
\end{eqnarray}
\normalsize
The second equality follows the fact that the probability of the outcome $\alpha$ is $p_{\alpha}=Tr((\id_2\otimes E_{\alpha})\rho_{q,MS})=\frac{1}{2}(p_{o,\alpha}+p_{e,\alpha})$. 

\section{Proof for the Entanglement Upper Bound}
\label{sec:proof}
Here we prove that $F_{avg,max}$ given in equation \ref{eq:fidavgmax} is the upper bound on the entanglement of the target qubits when the initial state of the MS is mixed state of the form \ref{eq:ThMixed}. We also show that the unitary gates $V_o^{\star}=e^{-i \pi/2 J_x}$ and $V_e^{\star}=\id^{\otimes N}$ and the set of the measurement operators $\{E_{\alpha}^{\star}\}=\{\Pi(m)\}$ saturate this upper bound. 



 The initial state of the MS can be expanded as,
\small
\begin{eqnarray}
\label{eq:roth1}
\rho_{\epsilon}&=&\left(\dfrac{\id+(1-\epsilon) \sigma_z}{2}\right)^{\otimes N}=\left((1-\dfrac{\epsilon}{2})\ketbra{0}{0}+(\dfrac{\epsilon}{2})\ketbra{1}{1}\right)^{\otimes N} \nonumber  \\
&=&\sum_{j=0}^{N} q^{N-j}(1-q)^{j} \sum_{i=1}^{\binom{N}{j}} P_i\left(\ketbra{0}{0}^{\otimes N-j}\otimes \ketbra{1}{1}^{\otimes j}\right)
\end{eqnarray}
\normalsize
where $q$ is defined as $q:=1-\frac{\epsilon}{2}$ and $P_i$ represents the permutation operator and the summation over $i$ is over all permutations. In a compact form, $\rho_{\epsilon}$ can be written as:
\begin{equation}
\label{eq:roth3}
\rho_{\epsilon}=\sum_{k=1}^{2^N}c_k \ketbra{\psi_k}{\psi_k}
\end{equation}
where $\{\ket{\psi_k}\}=\{\ket{0}^{\otimes N}, \ket{0}^{\otimes N-1}\otimes \ket{1},\ket{0}^{\otimes N-2}\otimes \ket{1}\otimes \ket{0}, ... ,\ket{1}^{\otimes N} \}$ is the orthonormal basis along the quantization axis ($Z$ axis) and the set of the corresponding coefficients is,
\small
\begin{eqnarray}
\label{eq:coef}
\{c_k\}&=&\{q^N, \binom{N}{1} \text{ times } q^{N-1}(1-q)  \nonumber \\
&&,\binom{N}{2} \text{ times } q^{N-2}(1-q)^2 , ..., (1-q)^N  \}.
\end{eqnarray}
\normalsize
With the assumption that $\epsilon < 1$, $q$ is bigger than $\frac{1}{2}$ and the above set of $\{c_k\}$ has a decreasing order.

Any unitary evolution rotates the orthonormal basis $\{\ket{\psi_k}\}$ to another orthonormal basis,
but it does not change the coefficients $c_k$; therefore the evolved states, $\rho_{o}$ and $\rho_{e}$, have similar expansions,
\begin{eqnarray}
\rho_{o}&=&V_{o}  \rho_{\epsilon} V_{o}^{\dagger}=\sum_{k=1}^{2^N}c_k \ketbra{\phi_{o,k}}{\phi_{o,k}} \nonumber \\
\rho_{e}&=&V_{e}  \rho_{\epsilon} V_{e}^{\dagger}=\sum_{k=1}^{2^N}c_k \ketbra{\phi_{e,k}}{\phi_{e,k}}
\end{eqnarray}
 where $\ket{\phi_{o,k}}=V_{o} \ket{\psi_k}$ and $\ket{\phi_{e,k}}=V_{e} \ket{\psi_k}$.  With a fixed POVM measurement, $\{E_{\alpha}\}$, the probability distribution of the measurement outcomes corresponding to the  state $\rho_o$ is,
\begin{eqnarray}
\label{eq:prob1}
p_{o,\alpha}&=&Tr\left(E_{\alpha}\rho_{o}\right)=\sum_{k=1}^{2^N} c_k Tr\left(E_{\alpha}\ketbra{\phi_{o,k}}{\phi_{o,k}}\right) \nonumber \\
 &=&\sum_{k=1}^{2^N} a_{o,\alpha k} c_k
\end{eqnarray}
 The variable $a_{o,\alpha k}:=Tr\left(E_{\alpha} \ketbra{\phi_{o,k}}{\phi_{o,k}}\right)$ has the following properties:
\small
\begin{eqnarray}
&&0\leq E_{\alpha} \leq \id  \longrightarrow  0 \leq a_{o,\alpha k} \leq 1 \nonumber \\
&&\sum_{k} a_{o,\alpha k} = Tr(E_{\alpha} \sum_k \ketbra{\phi_{o,k}}{\phi_{o,k}})=
Tr(E_{\alpha}) \nonumber \\
&&\sum_{\alpha} a_{o,\alpha k} = Tr((\sum_{\alpha} E_{\alpha})\ketbra{\phi_{o,k}}{\phi_{o,k}})
=1 \nonumber \\
&&\sum_{k,\alpha} a_{o,\alpha k}= \sum_{k} Tr( \id  \ketbra{\phi_{o,k}}{\phi_{o,k}})= \sum_{k} 1 = 2^N 
\end{eqnarray}
\normalsize
Similarly $p_{e,\alpha}=\sum_{k=1}^{2^N} a_{e,\alpha k} c_k$ with $a_{e,\alpha k}:=Tr\left(E_{\alpha} \ketbra{\phi_{e,k}}{\phi_{e,k}}\right)$ which has all the above properties.
Substituting the relations for $p_{o,\alpha}$ and $p_{e,\alpha}$ in equation \ref{eq:fidavg3} the average fidelity is,
\small
\begin{eqnarray}
F_{avg}&=&\frac{1}{2}\sum_{\alpha}  \max\left(\sum_{k=1}^{2^N} a_{o,\alpha k} c_k,\sum_{k=1}^{2^N} a_{e,\alpha k} c_k\right) \nonumber \\
&=&\frac{1}{2}\sum_{\alpha} \left( s_{\alpha}\sum_{k=1}^{2^N} a_{o,\alpha k} c_k+(1-s_{\alpha})\sum_{k=1}^{2^N} a_{e,\alpha k} c_k\right) \nonumber  \\
&=&\frac{1}{2}\sum_{k=1}^{2^N} \left(\left(\sum_{\alpha}s_{\alpha} a_{o,\alpha k}\right)+\left(\sum_{\alpha}(1-s_{\alpha}) a_{e,\alpha k}\right)\right) c_k  \nonumber \\
&=&\frac{1}{2}\sum_{k=1}^{2^N} \left(\beta_{o,k}+\beta_{e,k}\right) c_k=\frac{1}{2}\sum_{k=1}^{2^N} \beta_k c_k.
\end{eqnarray}
\normalsize
We introduce the coefficient $s_{\alpha}$, which acts as a switch; it is equal to $1$ if $p_{o,\alpha}\geq p_{e,\alpha}$ and equal to $0$ otherwise. We  also define $\beta_{o,k}:=\sum_{\alpha}s_{\alpha} a_{o,\alpha k}$, $\beta_{e,k}:=\sum_{\alpha}(1-s_{\alpha}) a_{e,\alpha k}$ and $\beta_{k}:=\beta_{o,k}+\beta_{e,k}$. Note that
 \begin{equation}
 \left. \begin{array}{ll}
 \beta_{o,k}=\sum_{\alpha}s_{\alpha} a_{o,\alpha k}\\
 \sum_{\alpha} a_{o,\alpha k}=1, \text{  } 0 \leq a_{o,\alpha k} \leq 1 \\
 s_{\alpha}=0,1 
 \end{array} \right\} \longrightarrow  0\leq \beta_{o,k} \leq 1
 \end{equation}
Similarly one can show that $0\leq \beta_{e,k} \leq 1$; and thus $0\leq \beta_{k} \leq 2$. Also note that the coefficients $\beta_k$ add up to $2^N$,
\begin{eqnarray}
\sum_{k}\beta_k &=& \sum_k \beta_{o,k}+\beta_{e,k}  \\
&=&\sum_k \left(\sum_{\alpha}s_{\alpha} a_{o,\alpha k} +\sum_{\alpha}(1-s_{\alpha}) a_{e,\alpha k}\right) \nonumber \\
&=&\sum_{\alpha} \left( s_{\alpha} \left(\sum_k a_{o,\alpha k} \right) +(1-s_{\alpha}) \left(\sum_k a_{e,\alpha k} \right) \right)\nonumber \\
&=&\sum_{\alpha} Tr(E_{\alpha}) (s_{\alpha}+1-s_{\alpha})=Tr (\sum_{\alpha} E_{\alpha})=2^N.\nonumber
\end{eqnarray}
\normalsize
As a summary,
\begin{equation}
\label{eq:FavgBeta}
F_{avg}=\frac{1}{2}\sum_k \beta_k c_k \text{   with   } 0\leq \beta_{k} \leq 2 \text{ , }  \sum_{k}^{2^N}\beta_k=2^N.
\end{equation}
 As mentioned before, the set of the coefficients $\{c_k\}$ is fixed by the initial state and with $q>\frac{1}{2}$ has the decreasing order displayed in equation \ref{eq:coef}. Thus in order to maximize $F_{avg}$ one should choose $\beta_k=2$ for half of $c_k$ with higher values and $\beta_{k}=0$ for the other half with lower values, if possible,
 i.e. 
 \begin{equation}
 \label{eq:alpha}
 \left\{ \begin{array}{ll}
 \beta_{k}=2 \hspace{2.2cm} 1\leq k \leq 2^{N-1}\\
  \beta_{k}=0 \hspace{1cm} 2^{N-1}+1 \leq k \leq 2^{N}
 \end{array} \right. .
  \end{equation}
This choice of $\beta_k$ results in the following maximum average fidelity and proves equation \ref{eq:fidavgmax}.
\begin{eqnarray}
\label{eq:fidavgmax3}
&&F_{avg,max}  \\
&&=\left\{ \begin{array}{ll}
\sum\limits_{l=0}^{\frac{N-1}{2}}(1-q)^{l} q^{N-l}\binom{N}{l} \hspace{3cm} \text{odd $N$} \\ 
 \sum\limits_{l=0}^{\frac{N}{2}-1}(1-q)^{l}q^{N-l}\binom{N}{l}+q^{\frac{N}{2}}(1-q)^{\frac{N}{2}}\binom{N}{\frac{N}{2}} \hspace{0.1cm} \text{even $N$} \end{array} \right. \nonumber \\
&&= \begin{cases}
B(\frac{N-1}{2};N,1-q)  &\text{odd $N$} \\
B(\frac{N}{2}-1;N,1-q)+\frac{1}{2}b(\frac{N}{2};N,1-q) &\text{even $N$}
\end{cases} \nonumber \\
&&=\begin{cases}
B(\frac{N-1}{2};N,\frac{\epsilon}{2})  &\text{odd $N$} \\
B(\frac{N}{2}-1;N,\frac{\epsilon}{2})+\frac{1}{2}b(\frac{N}{2};N,\frac{\epsilon}{2}) &\text{even $N$}  \nonumber
\end{cases} .
\end{eqnarray}
\normalsize
It is straight forward to show that the choices of $V_{e}^{\star}=\id^{\otimes N}$ and  $V_{o}^{\star}=\sigma_x^{\otimes N}$ and $\{E_{\alpha}^{\star}\}=\{\Pi_{\alpha}\}$ saturates the above upper bound for the average fidelity.
The probability distributions associated with the state $\rho_{e}^{\star}=V_e^{\star} . \rho_{\epsilon}.V_e^{\star\dagger}=\rho_{\epsilon}=(\frac{\id+(1-\epsilon) \sigma_z}{2})^{\otimes N}$ and  $\rho_{o}^{\star}=V_o^{\star} . \rho_{\epsilon}.V_o^{\star\dagger}=(\frac{\id-(1-\epsilon) \sigma_z}{2})^{\otimes N}$ are,
 \begin{eqnarray}
 \label{eq:P1s}
p_{o,\alpha}^{\star}&=&Tr(\Pi_{\alpha} \rho_{o}^{\star})=(1-q)^{\alpha} q^{N-\alpha}\binom{N}{\alpha}\\
p_{e,\alpha}^{\star}&=&Tr(\Pi_{\alpha}^{\star}\rho_{e}^{\star})=q^{\alpha} (1-q)^{N-\alpha}\binom{N}{\alpha}.\nonumber
 \end{eqnarray}
Since $q>\frac{1}{2}$, the $\max(p_{e,\alpha}^{\star},p_{o,\alpha}^{\star})$ is,
\begin{equation}
\max(p_{e,\alpha}^{\star},p_{o,\alpha}^{\star})=
\begin{cases} 
p_{o,\alpha}^{\star}=(1-q)^{\alpha} q^{N-\alpha}\binom{N}{\alpha} &\alpha\leq \frac{N}{2} \\
p_{e,\alpha}^{\star}=q^{\alpha} (1-q)^{N-\alpha}\binom{N}{\alpha} &\alpha > \frac{N}{2}
\end{cases}.
\end{equation}
Combining the above equality with the relation \ref{eq:fidavg3} for the average fidelity, results in the maximum average fidelity as given in equation \ref{eq:fidavgmax3}.


\bibliographystyle{apsrev4-1} 
\bibliography{references2}




\end{document}